\PassOptionsToPackage{unicode=true}{hyperref} 
\PassOptionsToPackage{hyphens}{url}
\PassOptionsToPackage{dvipsnames,svgnames*,x11names*}{xcolor}
\documentclass[a4paper,]{article}
\pdfoutput=1
\usepackage{lmodern}
\usepackage{amssymb,amsmath}
\usepackage{ifxetex,ifluatex}
\usepackage{fixltx2e} 
\ifnum 0\ifxetex 1\fi\ifluatex 1\fi=0 
  \usepackage[T1]{fontenc}
  \usepackage[utf8]{inputenc}
  \usepackage{textcomp} 
\else 
  \usepackage{unicode-math}
  \defaultfontfeatures{Ligatures=TeX,Scale=MatchLowercase}
\fi
\IfFileExists{upquote.sty}{\usepackage{upquote}}{}
\IfFileExists{microtype.sty}{%
\usepackage[]{microtype}
\UseMicrotypeSet[protrusion]{basicmath} 
}{}
\IfFileExists{parskip.sty}{%
\usepackage{parskip}
}{
\setlength{\parindent}{0pt}
\setlength{\parskip}{6pt plus 2pt minus 1pt}
}
\usepackage{xcolor}
\usepackage{hyperref}
\hypersetup{
            pdftitle={SAT solving techniques: a bibliography},
            colorlinks=true,
            linkcolor=magenta,
            citecolor=magenta,
            urlcolor=blue,
            breaklinks=true,
}
\ifx\paragraph\undefined\else
\let\oldparagraph\paragraph
\renewcommand{\paragraph}[1]{\oldparagraph{#1}\mbox{}}
\fi
\ifx\subparagraph\undefined\else
\let\oldsubparagraph\subparagraph
\renewcommand{\subparagraph}[1]{\oldsubparagraph{#1}\mbox{}}
\fi

\makeatletter
\def\fps@figure{htbp}
\makeatother

\usepackage[backref=true,backend=bibtex]{biblatex}
\title{SAT solving techniques: a bibliography}
\author{Louis Abraham\footnote{\href{mailto:louis.abraham@yahoo.fr}{\nolinkurl{louis.abraham@yahoo.fr}}}\\
\normalsize{École polytechnique}\\}
\date{\today}

\begin{document}
\maketitle
\begin{abstract}
We present a selective bibliography about efficient SAT solving, focused
on optimizations for the CDCL-based algorithms.
\end{abstract}

\hypertarget{introduction}{%
\section{Introduction}\label{introduction}}

SAT is one of the most famous NP-complete problems. However, today,
state-of-the-art solvers are able to solve industrial benchmarks with
more than one million variables and 10 million clauses despite using
exponential worst case algorithms \footnote{\url{https://www.princeton.edu/~chaff/zchaff.html}}.

The most performant solvers are based on two mechanisms: the boolean
constraint propagation (BCP) from the Davis--Putnam--Logemann--Loveland
(DPLL) algorithm, and its extension by conflict driven clause learning
(CDCL). The whole is commonly refered as CDCL.

The CDCL solvers are high feats of engineering that manage to tackle
very hard problems thanks to advanced heuristics. We gathered the main
articles that describe such optimizations.

\hypertarget{optimizations}{%
\section{Optimizations}\label{optimizations}}

We can distinguish three types of optimizations.

\hypertarget{algorithmic-optimizations}{%
\subsection{Algorithmic optimizations}\label{algorithmic-optimizations}}

Like any complex program, a CDCL solver relies on data structures.
Better data structures allow huge performance gains.

For example, detecting unit clauses (all but one literals are assigned
to \texttt{false}, the last one is free) and unsatisfied clauses is a
problem known as \textbf{boolean constraint propagation (BCP)}. One can
think of a simple counter-based algorithm, but now the best solvers use
lazy data structures like \emph{Head and Tail} or two watched literals
(2WL).\\
Another technique that seems performant, although not widely used, is
Early Conflict Detection Based BCP (ECDB).

The algorithm also uses a clause database, and might need priority
queues to implement the various heuristics described below.

\hypertarget{search-optimizations}{%
\subsection{Search optimizations}\label{search-optimizations}}

Like in any search program, the \textbf{order of the literal
assignments} is important. There are two types of strategies: static and
dynamic. Static strategies choose an order on the variables at the
beginning of the code while dynamic strategies make the order evolve
during the search.

The interest of dynamic strategies is reinforced by the ability to
\textbf{restart} the algorithm, that is delete all assignments of
variables.\\
A new execution of the algorithm may execute faster because of both the
updated order of assignment on the variables, and the learned clauses.

\hypertarget{clause-learning-and-deletion}{%
\subsection{Clause learning and
deletion}\label{clause-learning-and-deletion}}

When the current assignment does not satisfy a clause, one can deduce a
\textbf{conflicting clause} implied by the known clauses, but that
allows to find the conflict more easily if the situation occurs again.\\
While there are various ways to deduce such a conflicting clause, the
first unique implication point (1UIP) strategy is the most widely used.

However, learning too many clauses can also deteriorate the performance
because of memory overflow or an overuse of BCP.\\
Therefore, strategies arose to control the size of the clause database
and \textbf{forget clauses}.

\hypertarget{general-references}{%
\section{General references}\label{general-references}}

Knuth \autocite{knuth2015fascicle} gives a really broad vision of SAT
solving and discusses a broad range of topics like random SAT instances,
symmetry breaking or parallelism.

However, this bibliography is focused on the state-of-the-art solvers
that mostly use the CDCL paradigm.

A really good introduction with a lot of references is found in the blog
post \autocite{0aio2015sat}. It also explains why the lazy BCP data
structures cannot use pure literal elimination \autocite{44924}.

Biere et al. \autocite{biere2009conflict} formally explains the
principle of CDCL and lists the most discussed optimizations areas:
backtracking scheme, lazy data structures for BCP, restart strategies,
variable selection heuristics and clause deletion strategies.

Zhand and Malik \autocite{zhang2002quest}, although older, gives an idea
of what the state of the art was in 2002.

Some parts of a forthcoming textbook, \emph{Automated Reasoning---The
Art of Generic Problem Solving} by Weidenbach, are also available online
\autocite{weidenbach2018} .

Ryan \autocite{ryan2004efficient} gives detailled insights on the
various algorithms and heuristics, particularly for the BCP and the
decision strategy.

Fleury et al. \autocite{fleury2018verified} presents a totally
formalized algorithm using 2WL, thus it should be prefered to other
references because it proved the invariants of 2WL that are sometimes
insufficiently presented in other references.

\hypertarget{search-related-optimizations}{%
\section{Search-related
optimizations}\label{search-related-optimizations}}

\hypertarget{decision-heuristics}{%
\subsection{Decision heuristics}\label{decision-heuristics}}

In the absence of conflicts, the DPLL algorithms needs to make boolean
decisions on variables. There is no good strategy to choose between the
two possibilities (Knuth \autocite{knuth2015fascicle} suggests to
default to \texttt{false} for human-generated instances).\\
However, the choice of the variable to be decided is really important.

Today, most solvers use the VSIDS (Variable State Independent, Decaying
Sum) heuristic that updates priority scores for the literals.\\
Yi \autocite{yi2007effect} suggests VSIDS does not perform better than
random variable selection. However, this paper was neither published nor
quoted, according to Google Scholar.\\
Liang et al. \autocite{2015arXiv150608905L} uses the community
structures to explain why VSIDS works and improves it further.

Dershowitz et al. \autocite{dershowitz2005clause} compares three
families of decision heuristics: VSIDS, \emph{Berkmin} and their novel
Clause-Based Heuristic (CBH).\\
Biere and Fröhlich \autocite{biere2015evaluatingscoring} is the most
up-to-date reference: it compares 8 scoring schemes and designs a
generic queue data structure suitable for any scoring scheme. The ACIDS
(average conflict-index decision score) they introduct seems to be
competitive against EVSIDS (exponential VSIDS), the most commonly used
implementation of VSIDS, and VMTF (variable move-to-front) also named
\emph{Berkmin} because it was introduced by the \textsc{Berkmin} solver
\autocite{goldberg2002berkmin}. Furthermore, ACIDS does not involve any
parameters.

\hypertarget{restart-policies}{%
\subsection{Restart policies}\label{restart-policies}}

A solver can lose a lot of time exploring a barren part of the search
space.\\
The search is mostly determined by the decision heuristic and the
learned clauses. After a reasonable number of conflicts, the decision
literals on the trail will not be the maximal literals with respect to
the decision heuristic, and some useful clauses have been learnt.\\
Hence, restarting the solver by removing all assignments leads to a
difference in the execution because the decisions will be made in a
different order.

Huang \autocite{huang2007effect} compares several static restart
policies, and suggests the universally optimal policy for \emph{Las
Vegas} algorithms introduced by Luby \autocite{luby1993optimal} is the
best. However, as stated in the article, ``The theoretical relevance of
this property to clause learning remains an interesting question
though''.

Kautz et al. \autocite{kautz2002dynamic} studies dynamic restart
policies that improve the performances by 40\% to 65\% over Luby's.\\
Biere and Fröhlich \autocite{biere2015evaluatingrestart} compares the
performances of several restart policies in the state-of-the-art solver
\textsc{Glucose} \autocite{glucose}. They show their \texttt{static-256}
uniform policy performs similarly to Luby's policy. Furthermore, they
present the performances of various dynamic policies that outperform
static ones.

\hypertarget{cdcl-optimizations}{%
\section{CDCL optimizations}\label{cdcl-optimizations}}

\hypertarget{conflict-analysis}{%
\subsection{Conflict analysis}\label{conflict-analysis}}

Zhang et al. \autocite{zhang2001efficient} explains the basic principle
of learning: backtrack in a non-chronological way while learning
clauses. They compare several learning schemes and introduce the 1UIP
(first unique implication point) scheme. Their experiments showed that
the performance is not enhanced by decision schemes generating smaller
clauses ; and that the 1UIP scheme clearly outperforms the other
learning schemes.

\hypertarget{deleting-clauses}{%
\subsection{Deleting clauses}\label{deleting-clauses}}

The principle of CDCL is to learn useful clauses that will enhance the
quality of the search. The learned clauses are consequences of the
clauses given in the input, but some are more useful than others.\\
On the other hand, adding clauses has a cost for BCP, thus the need for
clause deletion.

Most strategies tend to keep smaller clauses and delete bigger ones more
aggressively because the BCP overload depends on the clause size.

Audemard and Simon \autocite{audemard2009predicting} compares strategies
based on clause activity (similar to VSIDS for variables) with their own
strategy based on Literals Blocks Distance (LBD). They detect special
clauses that are always kept independently of their size and named
``Glue Clauses''. Their algorithm was implemented in the
\textsc{Glucose} solver \autocite{glucose}.

Jabbour et al. \autocite{2014arXiv1402.1956J} introduces a simple
size-based randomized policy and proves it yields better results than
the LBD policy of \textsc{Glucose}.

\hypertarget{implementation-experiments}{%
\section{Implementation experiments}\label{implementation-experiments}}

Many solvers were implemented to test different optimizations.

Moskewicz et al. \autocite{moskewicz2001chaff} presents \textsc{Chaff},
a solver that brought a revolution in the world of SAT-solving by
introducing two major optimizations that we discussed: the 2WL BCP
algorithm and the VSIDS decision heuristic.\\
Eén and Sörensson \autocite{een2003extensible} explains the design of
\textsc{MiniSat} \autocite{minisat}, an implementation inspired by
\textsc{Chaff}. They also introduced the VSIDS activity for clauses.\\
Katebi et al. \autocite{katebi2011empirical} modified \textsc{MiniSat}
to rank the ``usefulness'' of its features. Thus, clause learning is the
most useful, followed by VSIDS. 2WL (compared to counter-based BCP) and
Luby's restart policy come after but are still responsible for major
performance improvements.\\
\textsc{Glucose} \autocite{glucose} is a modification of
\textsc{MiniSat} based on the alternative scoring scheme for the clause
learning mechanism presented in \autocite{audemard2009predicting}.

The \textsc{Berkmin} solver \autocite{goldberg2002berkmin} is mainly
known for the VMTF branching decision heuristic it used. It is mainly
based on the idea that recently deduced clauses are the most important
to satisfy.

The \textsc{MIRA} solver \autocites{lewis2004early}{lewis2005speedup}
implemented two novel optimizations: Implication Queue Sorting (IQS) and
Early Conflict Detection Based BCP (ECDB). It also combined the VSIDS
and VMTF branching heuristics. According to the benchmarks, it
outperformed the state-of-the-art version of \textsc{zChaff} that was
available at the time. However, the techniques it introduced were not
reused, and the article itself was not widely cited.\\
The details of ECDB are developed in Lewis et al.
\autocite{lewis2004early}, and cover an heuristic to first propagate the
implications that are the most likely to cause a conflict.

\hypertarget{conclusion}{%
\section{Conclusion}\label{conclusion}}

This overview of the evolution of SAT solving allows us to make some
observations. SAT solvers are a very empirical field and most techniques
are proved useful through experiments without being explained on a
theoretical point of view. The power of clause learning as a proof
system was only partially explained in 2004 by Beame et al.
\autocite{beame2004towards}, years after the advent of powerful CDCL
solvers.\\
It is difficult to conclude on the efficiency of a technique because the
performance can differ between the implementations and the hardware
evolution changed the way algorithms are executed.\\
The performance of a technique can also depend on other aspects of a
solver, thus explaining differences in the results.

Hopefully, the strong interest of the community allowed a
standardization of the benchmarks and the emergence of modular solvers
like \textsc{MiniSat} that can be easily modified to conduct meaningful
experiments.

\hypertarget{acknowledgments}{%
\section{Acknowledgments}\label{acknowledgments}}

We would like to thank Stéphane Graham-Lengrand for valuable discussions
and comments.

\nocite{*}
\printbibliography

\end{document}